\documentclass[]{spie}  

\usepackage{amsmath,amsfonts,amssymb}
\usepackage{graphicx}
\usepackage{hyperref}
\hypersetup{colorlinks,allcolors=blue}

\title{Scalable dataset acquisition for data-driven lensless imaging}

\author{Clara S. Hung}
\author{Leyla A. Kabuli}
\author{Vasilisa Ponomarenko}
\author{Laura Waller}
\affil{Department of Electrical Engineering and Computer Sciences\\ University of California, Berkeley, USA}

\authorinfo{Further author information: (Send correspondence to C.S.H.)\\C.S.H.: clarahung@berkeley.edu}

\begin{document} 
\maketitle

\begin{abstract}
Data-driven developments in lensless imaging, such as machine learning-based reconstruction algorithms, require large datasets. In this work, we introduce a data acquisition pipeline that can capture from multiple lensless imaging systems in parallel, under the same imaging conditions, and paired with computational ground truth registration. We provide an open-access 25,000 image dataset with two lensless imagers, a reproducible hardware setup, and open-source camera synchronization code. Experimental datasets from our system can enable data-driven developments in lensless imaging, such as machine learning-based reconstruction algorithms and end-to-end system design.
\end{abstract}


\section{INTRODUCTION}
\label{sec:intro}  

Lensless imagers are low-cost, compact computational cameras in which a traditional lens is replaced by a thin optical element (e.g. a phase mask) placed directly in front of the sensor. Unlike lensed cameras, which directly capture an image of the original scene in a one-to-one measurement, lensless imaging systems capture a one-to-many, or multiplexed, sensor measurement. This multiplexed measurement is usually modeled as a convolution between the scene and a point spread function (PSF), whose properties are determined by the optical system~\cite{RecentAdvances}. The original scene is reconstructed from the multiplexed measurement using an iterative or learned inverse algorithm, such as least squares with FISTA~\cite{FISTA}. 

Despite these advantages, the image quality of current lensless systems is not sufficient to compete with consumer cameras. An emerging body of work advocates for the use of machine learning-based reconstruction algorithms to enhance lensless image quality~\cite{Kristina, LenslessTransformer, Khan_2019_ICCV}. These methods are extremely data-intensive, requiring large datasets with tens of thousands of lensless measurements and their corresponding ground truth images to train the algorithms. Furthermore, recent work in lensless imaging and computational imaging is increasingly data-driven. In addition to machine learning-based reconstruction algorithms~\cite{Kristina, LenslessTransformer, Khan_2019_ICCV}, information-theoretic analyses~\cite{infotheory, infoanalysis, highdimimaging} and end-to-end system design~\cite{ericE2E, kellman} all rely on large training datasets. Experimental datasets are preferred over simulated data as they include the imaging non-idealities of real systems. 

In practice, acquiring experimental datasets presents several challenges: imagers require expertise to align and calibrate, and datasets need extensive automation and time to capture, leaving little flexibility for customization to unique research needs. To evaluate and fairly compare different lensless imaging systems, measurements must be acquired in parallel under the same imaging conditions. These experimental challenges mean few datasets are available~\cite{Kristina, LenslessTransformer, MWDN, FlatCamFace, FlatNet, Qian:24, ge2024LPSNet}, and of the available datasets, not all meet the current data demands. For example, with 10,000 images, the dataset from Khan et al.~\cite{FlatNet} is not large enough to train current state-of-the-art machine learning algorithms. Others are not publicly available~\cite{ge2024LPSNet, Qian:24}. Publicly available datasets large enough for such models~\cite{Kristina, FlatCamFace, MWDN, LenslessTransformer}, with 25,000 images, only contain data for a single phase mask, thus, they cannot be used to compare different lensless imaging systems. To address this data demand, we present an open-source dataset acquisition system that can acquire large, experimental datasets with high quality images and multiple lensless imagers in parallel. Our contribution meets all of the data demands for facilitating further data-driven developments in lensless imaging.

\section{Methods}
Our dataset acquisition pipeline consists of a hardware system for data acquisition and software framework for hardware control and computational processing. Measurements from multiple lensless imaging systems and a ground truth lensed camera are captured in parallel, as visualized in Fig.~\ref{fig:pipeline}, and computationally aligned to the ground truth image.

\subsection{Hardware}

\begin{figure} [ht]
   \begin{center}
   \begin{tabular}{c} \includegraphics[height=6cm]{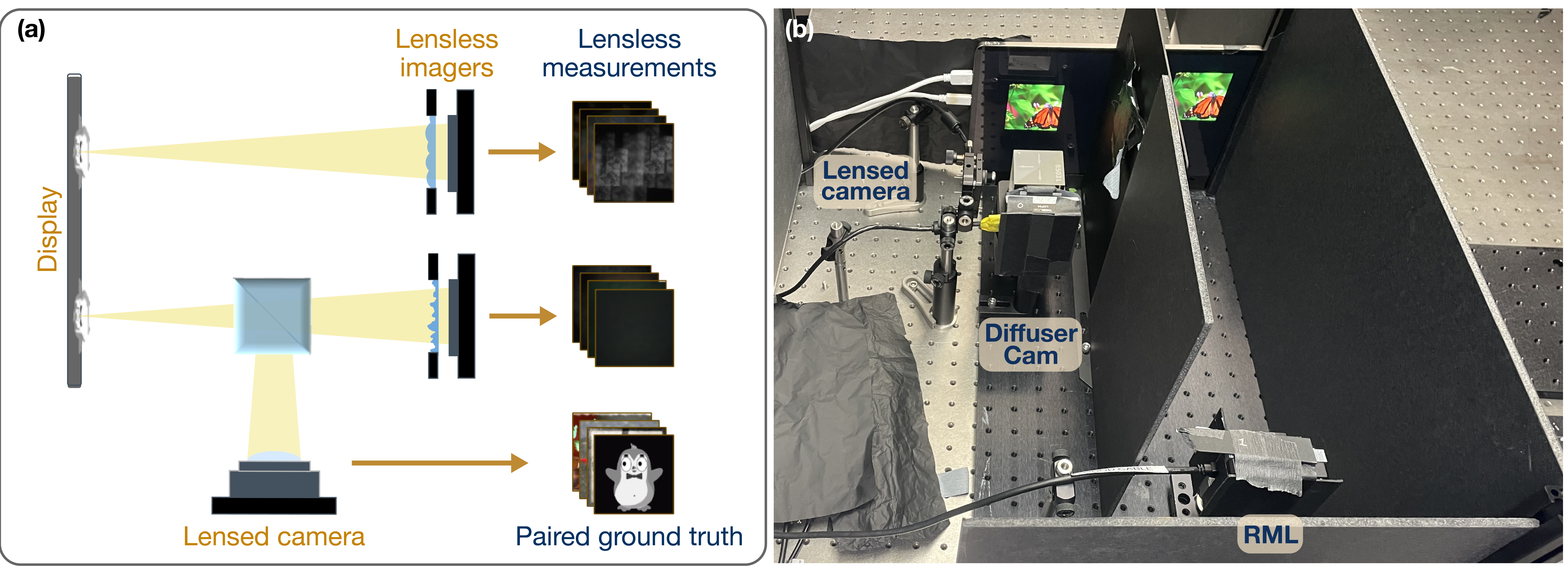}
   \end{tabular}
   \end{center}
   \caption[pipeline] 
   { \label{fig:pipeline} 
\textbf{Parallel dataset acquisition setup:} \textbf{(a)} A diagram of the hardware system. Images are displayed in parallel with two lensless imagers capturing lensless measurements and a lensed camera, which captures ground truth. \textbf{(b)} Experimental image of the setup, with two images displayed in parallel for simultaneous capture by all imagers.}
   \end{figure} 

We use a portable monitor (INNOCN 13A1F) to display images from the ground truth dataset~\cite{Mirflickr}, as shown in Fig.~\ref{fig:pipeline}. We demonstrate our pipeline’s capabilities by comparing two lensless imaging systems: one using a Gaussian diffuser (DiffuserCam\cite{DiffuserCam}) and another using a Random Multi-focal Lenslet (RML\cite{RML}) phase mask. The lensless imagers are built by aligning each imager’s phase mask in front of a board-level sensor (Basler daA1920-160uc). For DiffuserCam, we place the diffuser (Luminit 0.5°) 4.5~mm from the sensor and for the RML, the phase mask is 18~mm from the sensor, corresponding to distances with the sharpest PSFs. The imagers are mounted using a custom, 3D-printed mount and are aligned side by side, with 135~mm between them, so datasets from both imaging systems can be captured simultaneously (Fig.~\ref{fig:pipeline}a). Each phase mask includes an aperture to limit the PSF extent to the center of the sensor. The DiffuserCam and the RML capture lensless measurements, while a lensed camera (Basler daA1920-160uc Evetar S-mount, f=6~mm lens) images the same scene to acquire ground truth images. The lensed camera is aligned to a 50/50 split ratio beamsplitter (ThorLabs BS031) placed between the display and the DiffuserCam. We chose to split the lensed camera with the DiffuserCam instead of the RML as a result of the higher light throughput of the Gaussian diffuser. To control for stray light, we insert opaque black dividers between the RML and DiffuserCam and build an enclosure around the set up (see Fig. \ref{fig:pipeline}b).

\subsection{Software}
\begin{figure} [ht]
   \begin{center}
   \begin{tabular}{c} \includegraphics[height=7cm]{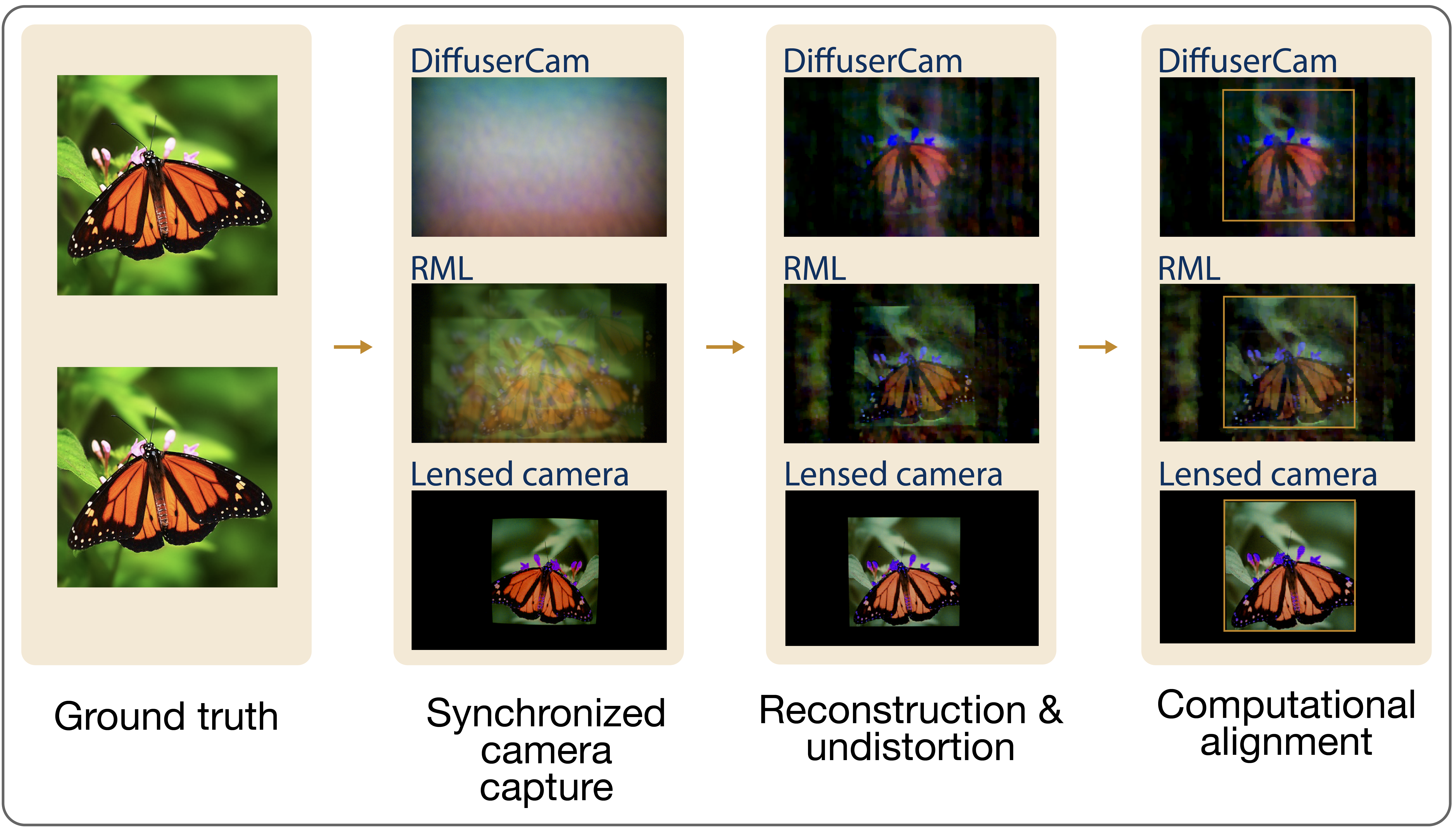}
   \end{tabular}
   \end{center}
   \caption[software] 
   { \label{fig:software} 
\textbf{Software pipeline for image and camera control.} The displayed images pass through each imaging system to generate measurements. Lensless measurements are computationally reconstructed while the ground truth lensed measurement is undistorted. Reconstructed images are computationally aligned to the undistorted ground truth image.}
   \end{figure} 
   
We developed a software package to automate image display and camera capture, as visualized in Fig.~\ref{fig:software} (\url{https://github.com/Waller-Lab/parallel-lensless-dataset}). The monitor is controlled with a script to automatically display images from the ground truth dataset in parallel, one for each lensless imager. For our 25,000 image dataset, we chose the open-source MIRFLICKR-25000 dataset~\cite{Mirflickr}. As the MIRFLICKR-25000 dataset images are of different dimensions, we crop each image to 300$\times$300 pixels before displaying them. For each image, the data acquisition of each camera is triggered synchronously, with a 200~ms delay between each camera and 500~ms delay between each image. This allows for the continuous acquisition of thousands of images. Additionally, images in the MIRFLICKR-25000 dataset have varying brightnesses, so we calibrated exposures for each of the imagers heuristically with respect to a reference image. For our dataset, we used one image from the MIRFLICKR-25000 dataset as the reference image and chose exposures to avoid over-saturation of the reference image measurement. 

For near real-time feedback during calibration and alignment, we use an image reconstruction script that reconstructs images using 200 iterations of FISTA~\cite{FISTA}. To correct for lens distortion in the ground truth captures, we calibrated the lensed camera by capturing calibration images of the OpenCV 11$\times$4 asymmetric circles grid~\cite{OpenCV}. To computationally align the reconstructed images with the undistorted ground truth images, we use the Kornia Python package~\cite{kornia} to learn a homography from the camera perspectives of each lensless imager to the lensed camera. This corrects small shifts that cannot be adjusted by optical alignment, ensuring pixel-to-pixel alignment of the images.

\subsection{System Calibration \& Analysis}
\begin{figure} [ht]
   \begin{center}
   \begin{tabular}{c}
   \includegraphics[height=7.5cm]{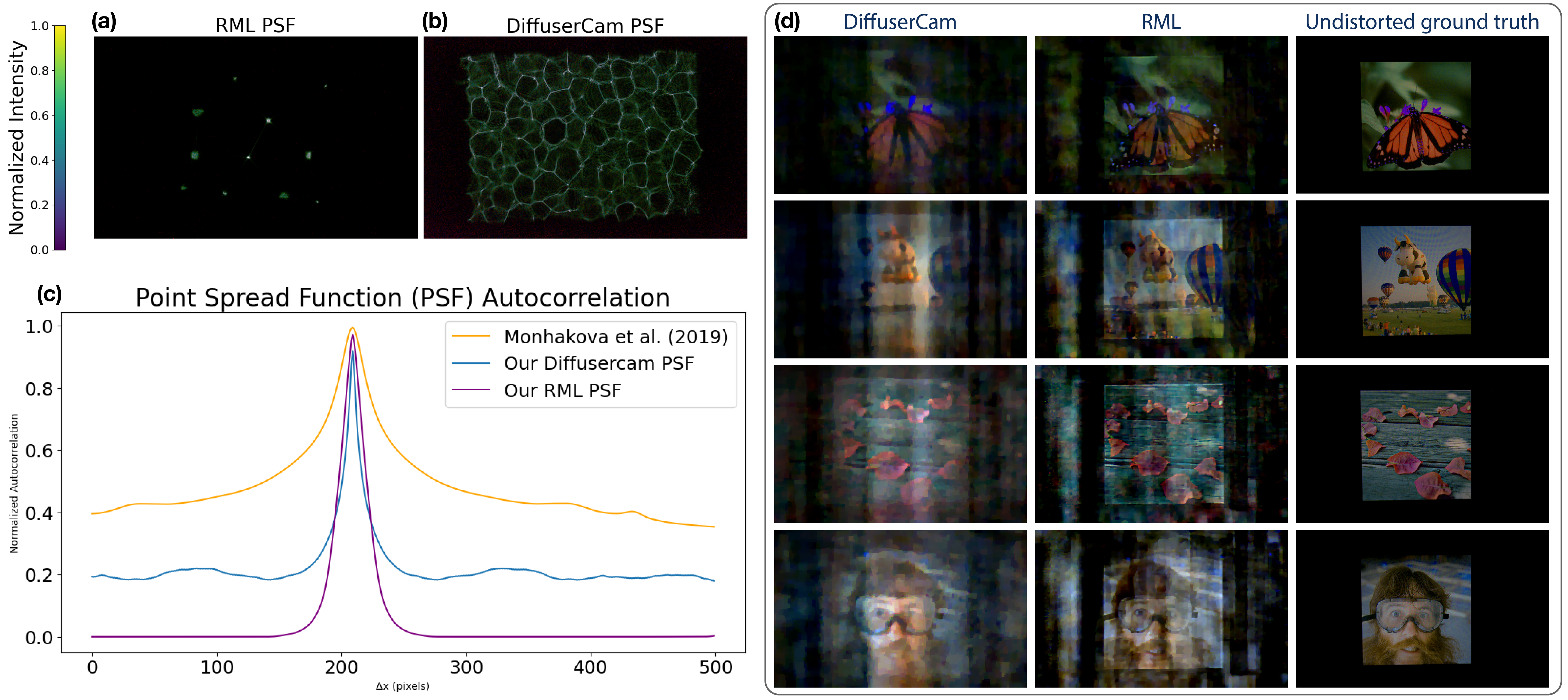}
   \end{tabular}
   \end{center}
   \caption[autocorr] 
   { \label{fig:autocorr} 
\textbf{System calibration PSFs}. \textbf{(a)} Our DiffuserCam point spread function (PSF). \textbf{(b)} Our Random Multi-focal Lenslet (RML) PSF. \textbf{(c)} An autocorrelation comparison between our DiffuserCam PSF, our RML PSF, and previous work that used a Gaussian diffuser Monhakova et al.~\cite{Kristina}. Both our DiffuserCam PSF (blue) and RML PSF (purple) have a sharper main lobe and lower sidelobes, indicating higher resolution and higher signal-to-noise ratio (SNR) compared to Monhakova et al.~\cite{Kristina}. \textbf{(d)} Sample reconstructions of DiffuserCam and RML measurements and undistorted ground truth images acquired using our system.}
   \end{figure}

Calibration PSFs for each lensless imager are captured by placing a point source at the imaging plane, 165~mm and 450~mm away from the phase mask for the DiffuserCam and RML, respectively. These distances were chosen for equal magnification between the DiffuserCam and RML's reconstructed images. To understand quality and resolution improvements compared to an existing DiffuserCam dataset~\cite{Kristina}, we compute autocorrelations of the PSFs. Resolution is determined by the autocorrelation peak width while sidelobes determine quality and SNR, approaching a delta function for an ideal system. Figure~\ref{fig:autocorr}c shows that our DiffuserCam and RML PSFs both have a narrower peak and lower sidelobes than Monhakova et al.~\cite{Kristina}, indicating our setup is able to capture measurements at a resolution and SNR that exceeds the quality of the current available state-of-the-art dataset~\cite{Kristina}. The RML autocorrelation has minimal sidelobes, improving SNR and image quality compared to the DiffuserCam (see Fig.~\ref{fig:autocorr}d). We present sample reconstructions in Fig.~\ref{fig:autocorr}d, using 200 iterations of FISTA~\cite{FISTA}, of our DiffuserCam and RML measurements and the ground truth images corrected for lens distortion. 

\section{Discussion}
Our dataset acquisition pipeline is designed to be reproducible, with a modular hardware system and open-source software package. Future work can acquire datasets larger than 25,000 images and extend the system to acquire 3D and microscopy-scale datasets. Using an OLED display (e.g. M4 iPad Pro) could yield further image quality and color fidelity improvements, as an OLED display has true black levels. We can easily adapt our framework to capture video datasets, which could be applied to recent developments in space-time reconstruction methods~\cite{tiff}.

\section{Conclusion}
We provide a framework for parallel lensless dataset acquisition and an open-access, 25,000 image dataset, both of which can be accessed at: \url{https://waller-lab.github.io/parallel-lensless-dataset}. Datasets acquired from our system include measurements from two lensless imagers under identical imaging conditions and with paired ground truth, making it possible to evaluate and compare lensless imaging systems. Although we demonstrate our system using the RML and DiffuserCam, the modular hardware system can be extended to use lensless imagers with different phase masks. 

Our contribution enables a quantitative understanding of lensless imaging system performance for use in data-driven applications, such as machine learning-based image reconstruction algorithms~\cite{Kristina, LenslessTransformer}, information-theoretic analyses~\cite{infotheory, infoanalysis, highdimimaging} and end-to-end system design~\cite{ericE2E, kellman}. Applying current state-of-the-art machine learning models, e.g. transformers, to image reconstruction shows promise for improving lensless image quality, and datasets from our system can be used to train such networks. 

\acknowledgments
The authors thank Eric Bezzam, Dekel Galor, and Neerja Aggarwal for helpful discussions and Eric Markley for 3D printing assistance. The authors were supported by the Chan-Zuckerberg BiohubSF. C.S.H. was supported by STROBE SURS, UC Berkeley L\&S SURF, and the Astronaut Scholarship Foundation. L.A.K was supported by the National Science Foundation Graduate Research Fellowship Program under Grant DGE 2146752.
 
\bibliography{main} 
\bibliographystyle{spiebib} 

\end{document}